\documentclass[twocolumn,apjl]{openjournal}

\newcommand{\Kepler}{{\it Kepler}}

\newcommand{\vespa}{\texttt{vespa}}

\newcommand{\be}{\begin{equation}}
\newcommand{\ee}{\end{equation}}

\usepackage{xcolor}
\definecolor{my_color}{HTML}{3a18b1}

\definecolor{new_color}{HTML}{CF0000}

\definecolor{new_black}{HTML}{000000}
\newcommand*{\bedit}{\textcolor{new_black}}

\newcommand{\msun}{M$_\odot$}
\newcommand{\rsun}{R$_\odot$}

\newcommand{\rearth}{R$_\oplus$}

\usepackage{xcolor}
\usepackage{hyperref}
\usepackage{breakurl}
\usepackage{amsmath}
\usepackage{graphicx}
\usepackage{verbatim}
\usepackage{booktabs}
\usepackage{orcidlink}

\hypersetup{
    colorlinks,
    linkcolor={red!50!black},
    citecolor={blue!50!black},
    urlcolor={blue!80!black}
}

\slugcomment{}

\shorttitle{New False Positives from the K2 mission}
\shortauthors{Lehmann \& Vanderburg}

\begin{document}



\title{Ephemeris Matching Reveals False Positive Validated and Candidate Planets from the K2 Mission}

\author{Drake A. Lehmann\altaffilmark{1,$\dagger$}\orcidlink{0009-0009-7983-801X}, Andrew Vanderburg\altaffilmark{2,$\star$}\orcidlink{0000-0001-7246-5438}}
%

\altaffiltext{1}{Department of Astronomy, University of Wisconsin-Madison, Madison, WI 53706, USA}
\altaffiltext{2}{Department of Physics and Kavli Institute for Astrophysics and Space Research, Massachusetts Institute of Technology, Cambridge, MA 02139, USA}
\altaffiltext{$\dagger$}{\url{dalehmann@wisc.edu}}
\altaffiltext{$\star$}{\url{andrewv@mit.edu}}


\begin{abstract}

Data from the \Kepler\ space telescope have led to the discovery of thousands of planet candidates. Most of these candidates are likely to be real exoplanets, but a significant number of false positives still contaminate the sample, especially in candidate lists from the K2 mission. Identifying and rejecting the false positives lurking in the planet candidates sample is important for prioritizing follow-up resources and measuring the most accurate population statistics. Here, we identify false positives in the K2 planet candidate sample using a technique called ``ephemeris matching," in which we compare the period and transit time of different signals. When signals from different stars show the same period and time of transit, we can conclude that at least one of the two signals is contamination. We identify \bedit{43} false positives among published K2 planet candidates (nearly 2\% of the complete list), one of which (K2-256 b) was previously validated as genuine exoplanet. This work increases the reliability of the K2 planet sample and helps boost confidence in the surviving planet candidates.    

\end{abstract}


\section{Introduction}

Six years after it exhausted its fuel reserves and ended operations, the \Kepler\ space telescope remains history's most prolific exoplanet detecting instrument. Most of these discoveries came from the data collected during its first four years of observations (2009-2013), when \Kepler\ pointed with extreme stability at a 110 square degree region of sky in the Northern constellations of Cygnus and Lyra. During that time, \Kepler\ observed over 200,000 stars and detected several thousands of planet candidates \citep{Thompson2018ApJS}, most of which are believed to be genuine exoplanets \citep[e.g.,][]{Latham2011, morton12, Morton2016ApJ}. After the failure of two of its four reaction wheels (gyroscope-like devices used to keep the spacecraft pointed steadily) in 2013, \Kepler\ was unable to continue observations of its primary field. NASA therefore transitioned to operating the K2 mission, in which \Kepler\ observed new fields along the ecliptic plane with poorer pointing precision \citep{howell}. \Kepler\ continued operating in its K2 mode until it exhausted its fuel reserves in 2018.  

K2 experienced repeated pointing excursions of a few arcseconds on timescales of $\approx$6 hours which caused significant systematic errors in the mission's light curves. As a result, the pipeline responsible for producing light curves in \Kepler's original mission was unable to produce light curves for the first year and a half of the K2 mission, and the K2 mission team never conducted an official transit search. Instead, a number of pipelines developed by members of the community were the primary source of light curves and planet candidates during the K2 era. These efforts were highly successful, resulting in the discovery of over 1000 planet candidates and over 500 confirmed planets \citep{akeson}.

\begin{deluxetable*}{ccccccccc}\label{dat_van}
\centering
\tabletypesize{\small}
\tablewidth{0pt}
\tablecaption{Full list of periodic transit/eclipse signals in K2\label{dat_van}}
\tablehead{\colhead{EPIC ID} & \colhead{RA} & \colhead{DEC} & \colhead{\Kepler} & \colhead{Period} & \colhead{$t_0$} & \colhead{Duration} & \colhead{Campaign} & \colhead{Label}\\\colhead{} & \colhead{(deg)} & \colhead{(deg)} & \colhead{magnitude} & \colhead{(days)} & \colhead{(BJD-2454833)} & \colhead{(hours)} & \colhead{} & \colhead{}}
\startdata
201126503 & 175.9384 & -5.873   & 17.275 & 1.19487  & 1977.376 & 1.793255 & 1        & C     \\
201146489 & 173.4322 & -5.37874 & 13.216 & 21.42847 & 1989.586 & 6.004302 & 1        & E     \\
201155177 & 176.6657 & -5.17191 & 14.632 & 6.686831 & 1981.678 & 2.891901 & 1        & C     \\
201158453 & 173.9807 & -5.09388 & 14.69  & 7.077647 & 1980.479 & 5.152039 & 1        & E     \\
201160323 & 175.7146 & -5.04876 & 18.238 & 22.27021 & 1978.497 & 4.283882 & 1        & E     \\
201161715 & 174.6153 & -5.01563 & 14.652 & 10.4953  & 1986.425 & 18.44555 & 1        & E     \\
201173390 & 169.1629 & -4.73515 & 12.964 & 16.99538 & 1983.43  & 12.44053 & 1        & E     \\
201182911 & 172.7104 & -4.50761 & 15.516 & 0.996561 & 1978.09  & 2.30753  & 1        & E     \\
201184068 & 173.4971 & -4.47994 & 14.135 & 0.794282 & 1977.576 & 2.608427 & 1        & E     \\
201197348 & 177.095  & -4.16439 & 15.135 & 14.90997 & 1992.115 & 2.472511 & 1        & C   \\
... &   &  &  &  &  &  &       & 
 
\enddata
\tablecomments{The full table is available in the arXiv source. Label refers to the classification of the signal (either C for ``candidate'' or E for ``eclipsing binary.'')}
\end{deluxetable*}

A consequence of the lack of an official K2 transit search is that the K2 planet candidate catalog is highly nonuniform. Different teams used different methods to produce light curves, identify candidates, and vet the results to pick out false positive signals. Since many analyses did not use all available tools developed during the original \Kepler\ mission to identify false positives, it is likely that a number of false positives remain within the list of K2 planet candidates.

In this work, we present a uniform search the list of known planet candidates from the K2 mission to identify false positives using a technique known as ephemeris matching \citep{Coughlin2014AJ}. This method entails a meticulous comparison of the periods and transit times (epochs) of planet candidates observed in two distinct and unrelated stars. If these values match, we can confidently assert that at least one of the candidates is a false positive due to contamination from the other.

A number of physical effects in in the \Kepler\ detector and optics the can cause different stars to show the same transit signal. In particular, the main physical effects causing ephemeris-matched signals are believed to be: 

\begin{enumerate}
\item \textit{Direct PRF contamination}: light from one star which scatters to nearby pixels on the CCD module, contaminating other stars in the vicinity. 

\item \textit{Antipodal reflection}: ghost images of extremely bright stars can appear across the focal plane due to light reflecting off of the telescope's correcting lenses, contaminating any stars directly across the focal plane from the source.

\item \textit{Column anomaly}: light from a star can contaminate other stars that happen to lie in the same CCD column (likely as a result of charge transfer inefficiency when reading out the detector \bedit{or imperfectly corrected CCD readout smear}).

\item \textit{CCD crosstalk}: a physical effect where wires that are bundled together can induce a ghost signal in each other. Results in a ``star" in the same pixel location on each of the outputs. If there happens to be a real star there, it is contaminated.

\end{enumerate}

In this paper, we conduct ephemeris matching on the K2 planet candidate catalog, using lists of candidates obtained from both the NASA Exoplanet Archive \citep{akeson} and the list of planet candidates and eclipsing binaries presented by \citet{Dattilo2019} for their training set. We successfully identify \bedit{43} false positives among these planet candidates, alongside one false positive-validated planet. Our paper is structured as follows: Section \ref{observations} provides an account of the observations we used as inputs to our analysis. Section \ref{analysis} details our methodology and the analytical processes applied to our data to perform ephemeris matching. Section \ref{results} delves into our findings in detail, and Section \ref{discussion} discusses the implications of our work and concludes.

\section{Observations}\label{observations}

All of the observations used in this paper come from the K2 mission. K2 conducted observations in approximately 80-day observing ``campaigns," during which the telescope was directed toward a different region of the sky on the ecliptic plane. Each campaign typically involved observations of around 30,000 celestial objects (mostly stars, but also some galaxies, solar system objects, or other astrophysical phenomena). After the completion of each campaign, the data were transmitted back to Earth, calibrated by the \Kepler\ pipeline, and released to the public. At this point, numerous teams downloaded the data in various stages of calibration, ranging from raw pixels \citep[e.g.,][]{Yu2018}, to calibrated pixels \bedit{\citep[e.g.,][]{Vanderburg2016, Luger2016AJ}}, to extracted light curves \citep[e.g.,][]{Pope2016}. These groups then performed their own analysis, produced light curves, removed systematic effects from the spacecraft's unstable pointing, searched for transits, triaged and vetted the signals, and published their results. 

Our ephemeris matching analysis requires two inputs: the list of planet candidates to be searched for false positives, and a comprehensive list of all periodic eclipse signals in the dataset (including planet candidates, but also eclipsing binary stars). In this section, we describe in more detail the source of these two inputs. 

\subsection{K2 Planet Candidates}

We compile our list of published planet candidates to search for false positives from two sources: the NASA Exoplanet Archive, and the compiled list of labeled periodic signals from \citet{Dattilo2019}. Since the beginning of the K2 mission, the NASA exoplanet archive has compiled the planet candidates reported in catalog papers from members of the community \bedit{\citep[e.g.,][]{Adams2016, Barros2016, Crossfield2016, Osborn2016MNRAS, Pope2016, Vanderburg2016, Livingston2018, Petigura2018, Yu2018, Kruse2019, Zink2021}}. We downloaded the K2 planet candidate list on June 30, 2022.

In addition to the planet candidates from the NASA Exoplanet Archive, we included the planet candidates reported by \citet{Dattilo2019}. \citet{Dattilo2019} identified numerous planet candidates (including many not included in the NASA Exoplanet Archive lists) and used them to train a neural network classifier for identifying viable planet candidates from K2. We include the objects labeled as candidates from their combined training/testing/validation set in our list. 

In total, we tested 2859 signals from \citet{Dattilo2019} and 3517 signals from  the NASA Exoplanet Archive. Both of these lists include duplicates of the same signal identified in different campaigns or by different authors, so the total number of planet candidates is somewhat lower than these totals. 

\subsection{Full List of Periodic Eclipse Signals}

There is no compiled list of all periodic eclipse signals (including both planet candidates and eclipsing binaries) from the K2 mission. We therefore compiled our own list. We base our list on the training/validation/test sets from \citet{Dattilo2019}, taking only the signals labeled as either planet candidates or eclipsing binaries. However, \citet{Dattilo2019} did not perform their analysis on the full set of K2 observations, so we supplemented their list to include some of the missing campaigns. In our analysis, we used all campaigns except for Campaign 0 (short campaign, few targets, few candidates), Campaign 9 (primarily a microlensing campaign, very few non-microlensing targets), and Campaign 19 (only one week of good data; see  \citealt{Incha2023MNRAS}).

Between 2014 and 2019, one of us (AV) routinely processed light curves from the K2 mission. Upon the completion of each campaign, the data were transmitted back to Earth, calibrated by the \Kepler\ pipeline, and released to the public. Once the data were released, we downloaded and analyzed them following the process outlined by \citet{Dattilo2019}. In particular, starting with the calibrated pixel files, we extracted light curves as described \citet{vj14} and \citet{Vanderburg2016}, searched for transits, performed triage on the detected signals (a quick by-eye sorting of the signals to separate planet candidates and eclipsing binaries from instrumental artifacts). These triaged light curves formed the basis\footnote{We note that \citet{Dattilo2019} performed additional vetting to remove additional false positive eclipsing signals from their list of candidates; we do not do the same because we only care about the presence of eclipses and periodicity for our analysis, not whether any given signal is likely a planet or an eclipsing binary.} for the dataset used by \citet{Dattilo2019} to train and test their neural network, but they did not use data from Campaigns 11, 17, and 18. We therefore supplement the \citet{Dattilo2019} dataset with our own lists of triaged eclipsing signals for these campaigns. In total, we identified 12322 periodic eclipsing signals for use in ephemeris matching. Our compiled list of eclipse-like signals is given in Table \ref{dat_van}.

\section{Analysis}\label{analysis}

\begin{figure*}[t!]  
    \centering
    \includegraphics[width=0.98\textwidth]{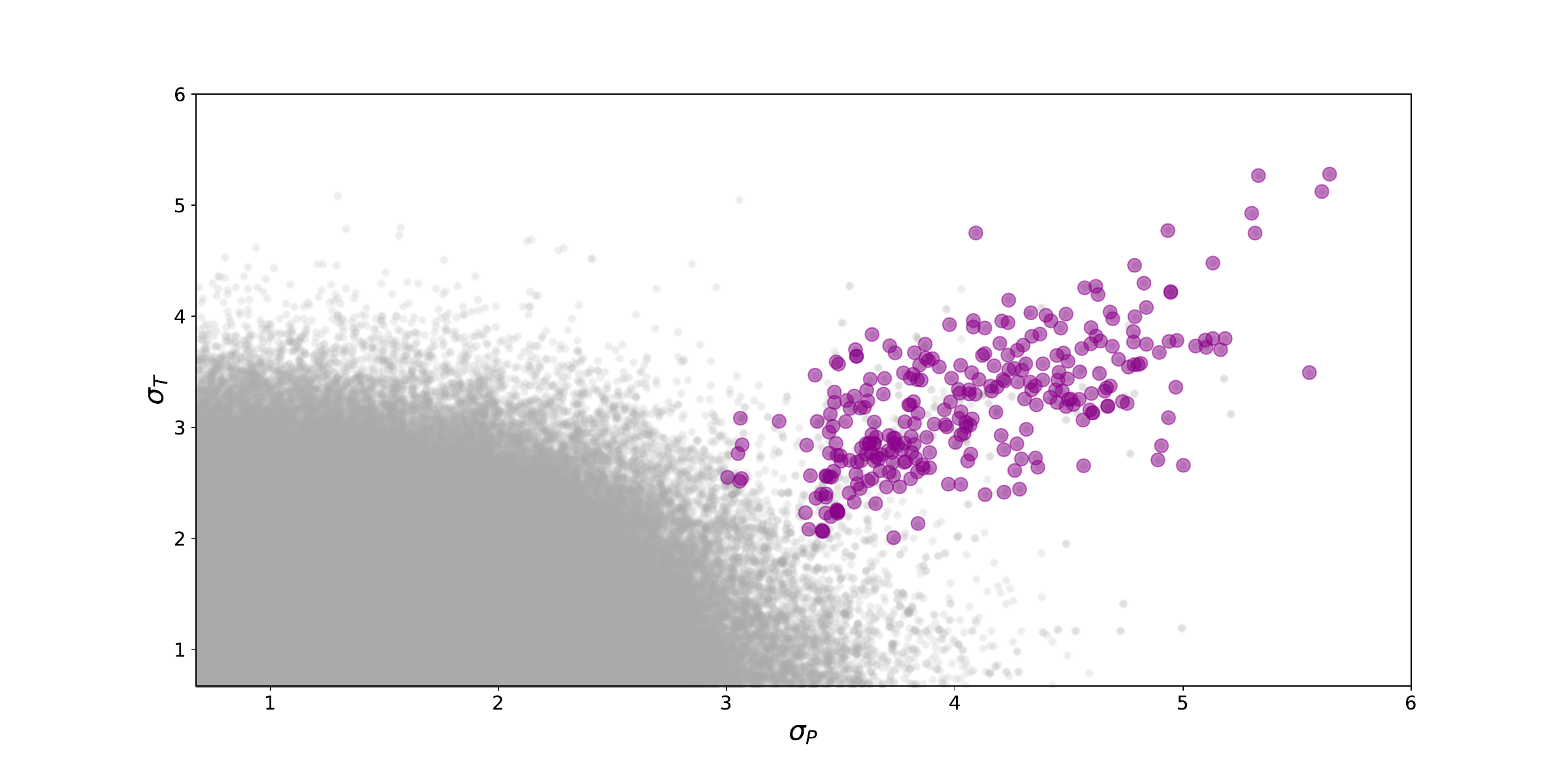}
    \caption{Significance values for all pairs of matched signals from the K2 mission (grey) and likely ephemeris matches (purple). The Y axis shows $\sigma_T$, the approximate probability of a chance match in transit time (given in terms of Gaussian standard deviations), while the X axis shows $\sigma_P$, the corresponding value for orbital period. A higher value of $\sigma_P$ indicates a closer resemblance in the periods of two planet candidates, while a larger $\sigma_T$ value indicates a greater similarity in transit times. Using criteria described in Section \ref{analysis}, we determined which of the grey points are likely true matches, and show these likely false positives in purple. \bedit{We note that there are a number of highly significant matches that failed at least one of the criteria described in Section \ref{analysis}; many of these cases are when the same star matched to its own signal. A few of the highly significant matches are cases of likely contamination that were rejected as false positives based on the criteria in Section \ref{analysis}. This is an indication that these criteria may be too strict, so future work could focus on modifying these criteria (that were originally devised for Kepler data) to better identify false positives in K2 data.}} 
    \label{Figure 1}
\end{figure*}

After collecting the lists of planet candidates and the lists of periodic eclipsing signals in K2, we performed ephemeris matching closely following the methodology of \citet{Coughlin2014AJ}. We started by collecting the various information for each signal needed for the analysis. In particular, our analysis required:
\begin{enumerate}
    \item An identifier for each star (we used the Ecliptic Plane Input Catalog, or EPIC; \citealt{Huber2016ApJS}). 
    \item Basic information about the star, including its right ascension,  declination, and \Kepler-band magnitude (also from the EPIC). 
    \item Information about the signal, including its period, time of eclipse/transit, eclipse/transit duration, and the campaign in which the signal was detected (from either the Exoplanet Archive, \citealt{Dattilo2019}, or our own lists). 
\end{enumerate}
Given these inputs, we then used the \texttt{K2FOV} package \citep{k2fov} to determine where each target fell on the \Kepler\ detectors (that is, on which CCD row, column, and module number the target was observed). 




To identify signals with matching ephemerides, we performed the following steps:

\begin{enumerate}
    \item We started by splitting up the signals in both lists by the campaign in which they were observed, in order to perform our matching analysis on a campaign-by-campaign basis.  Generally, we expect that any contaminating signals will come from other stars in the same part of the sky. Therefore, because K2 campaigns were spread all across the sky, there should be relatively few examples of ephemeris matches between stars observed in different campaigns. Moreover, restricting our analysis to one campaign significantly decreased the computational expense of identifying ephemeris matches (since the number of possible matching pairs scales with the number of candidates times the number of all eclipse/transit signals). 
    
    \item We found signals with periods and times of eclipse/transit so similar to one another that the match is statistically unlikely to be the result of random chance. We did this by closely following \citet{Coughlin2014AJ}. We calculated the fractional difference between the periods ($\Delta P$) and transit/eclipse times ($\Delta T$) for every possible pair of targets using the following equations: 

\begin{equation}
    \Delta P = \frac{P_A - P_B}{P_A}
\end{equation}

\begin{equation}
    \Delta T = \frac{T_A - T_B}{P_A}
\end{equation}

\noindent Where $P_A$ and $P_B$ are the periods detected on a pair of targets A and B, and $T_A$ and $T_B$ are the transit/eclipse times of those same targets. Next, we modify these fractional differences to find $\Delta P'$ and $\Delta T'$:

\begin{equation}
    \Delta P' = |\Delta P - {\rm int}(\Delta P)|
\end{equation}

\begin{equation}
    \Delta T' = |\Delta T - {\rm int}(\Delta T)|
\end{equation}

where the `int' function rounds the \bedit{fractional differences} to the nearest integer. This manipulation guarantees that two objects with the \bedit{nearly identical} periods and epochs will have $\Delta P'$ and $\Delta T'$ both equal to zero. The closeness to zero for these values determines the similarity between their periods and epochs. Again following \citet{Coughlin2014AJ}, we quantify this similarity in terms of Gaussian standard deviations $\sigma_P$ and $\sigma_T$: 

\begin{equation}
    \sigma_P = \sqrt{2} \cdot {\rm erfcinv}(\Delta P')
\end{equation}
\begin{equation}
    \sigma_T = \sqrt{2} \cdot {\rm erfcinv}(\Delta T')
\end{equation}

\noindent where the `erfcinv' function is the inverse complementary error function. Like \citet{Coughlin2014AJ}, we consider pairs of signals with $\sigma_P>3$ and  $\sigma_T>2$ to have matching ephemerides\footnote{\bedit{We note that these cutoffs are equivalent to transit times matching to within 5\% of an orbital period, and a fractional period difference of about 0.3\%}}. We show the $\sigma_P$ and  $\sigma_T$ values for all pairs of objects as grey dots in Figure \ref{Figure 1}. 


\item Next, we remove cases where a star matches with itself (that is, target A and target B happen to be the same star), since this is not a situation where two stars share the same signal. We do this by identifying cases where both target A and B have the same EPIC ID and removing them from the list.

\item As noted by \citet{Coughlin2014AJ}, given the large number of pairs of objects we are searching and the uncertainties in measured periods and transit/eclipse times, it is possible for two unrelated signals to match due to pure statistical chance. We therefore identify which of the signals with high values of $\sigma_P$ and  $\sigma_T$ are likely due to contamination by calculating heuristic criteria found by \citet{Coughlin2014AJ}. These criteria roughly determine which (if any) of the various physical processes that cause ephemeris matches can cause each potential match. First, we calculate $d_{max}$, which determines whether direct PRF contamination can likely cause an ephemeris match: 

\begin{equation}
    d_{max}(\arcsec) = 50 \cdot \sqrt{10^6 \cdot 10^{-0.4 \cdot m_{kep}} + 1}
\end{equation}

\noindent where $m_{kep}$ is the \Kepler-band magnitude of the brighter star, and where the units of $d_{max}$ are in arcseconds. Typically, direct PRF contamination is possible for stars up to about 50$\arcsec$ apart, but this distance increases with star's brightness. We then calculated the great-circle distance between each pair of targets. For each pair of targets with $\sigma_P>3$ and  $\sigma_T>2$ separated by a distance less than $d_{max}$, we considered the ephemeris match to be likely caused by contamination, indicating that at least one of the signals is likely a false positive. We show these objects as purple dots in Figure \ref{Figure 1}. 

\item Next, we identified pairs of stars that could have ephemeris matches due to column anomaly. Like \citet{Coughlin2014AJ}, we checked if each pair of stars was on the same CCD module and be situated within 10 CCD columns of each other or 10 CCD rows of each other. We consider pairs that satisfied both this criterion and had $\sigma_P>3$ and  $\sigma_T>2$ to be likely matches. We illustrate these matches in Figure \ref{Figure 1} in purple as well. 

\item Finally, we identified pairs of stars that could have ephemeris matches due to antipodal reflection. Again, following \citet{Coughlin2014AJ}, we checked whether each pair of stars were located directly across the \Kepler\ field of view from one another by \bedit{inverting} the coordinates across \Kepler's boresight position. If the two stars were within 50 arcseconds of each other's antipode and also satisfied $\sigma_P>3$ and  $\sigma_T>2$, we considered them to be likely matches and showed them in Figure \ref{Figure 1} in purple as well. 

\end{enumerate}

\begin{figure}[t!]
    \centering
    \includegraphics[width=0.49\textwidth]{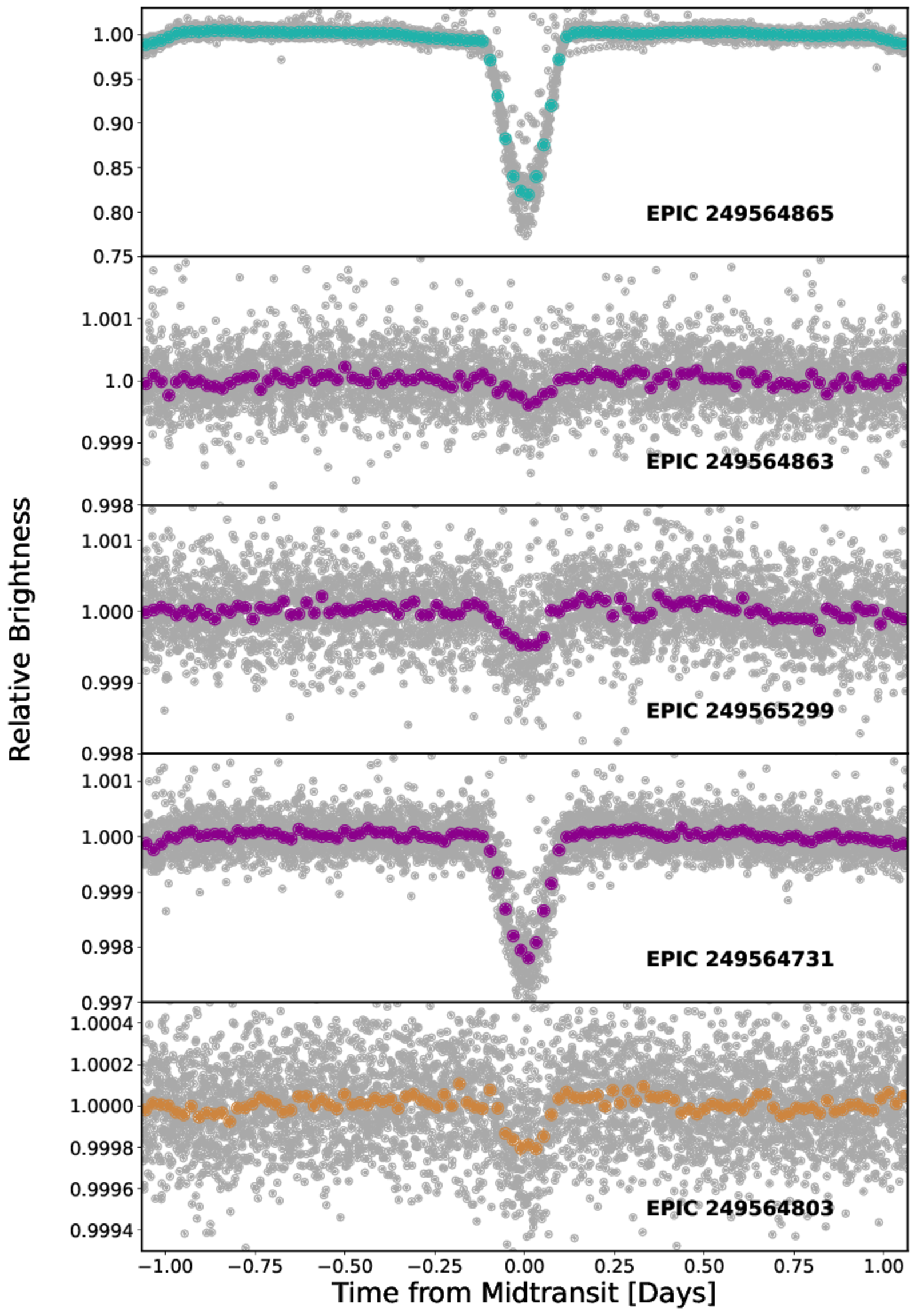} 
    \caption{Light curves of stars with matching ephemerides. In each, grey points are individual \Kepler\ long cadence measurements, while colored points are averages in phase. Here, we identify a family of 5 signals that all show the same period and transit/eclipse time, indicating a common source. Visual inspection shows that beyond having precisely the same period and transit time, the shapes and durations of the transit/eclipse signals are consistent. The stars hosting these light curves are all located nearby in sky (see Figure \ref{Figure 2}). We identify the light curve in the top panel (EPIC 249564865) as the likely source of all of the other signals, because of its large depth and the star's bright \Kepler-band magnitude ($K_p =$10.9). The light curves are colored based on the physical process causing the contamination. EPIC 249564865 is the likely source (shown with averaged points in teal). The stars with purple average points likely are contamination via direct PRF overlap. EPIC 249564803, with its orange averaged points, is likely contamination from column anomaly. }
    \label{Figure 3}
\end{figure}

After identifying signals with potentially matching ephemerides, we inspected each pair of signals to determine whether the ephemeris match implied that either of the two signals was a false positive planet candidate. We created diagnostic plots (similar to Figure \ref{Figure 3}) with the phase-folded light curves of the two signals to see whether their shapes were morphologically similar, and we inspected the \Kepler\ pixel images and photometric apertures of each light curve to determine which star was the origin of the signal in cases of direct PRF overlap. \bedit{The full set of diagnostic plots are available for download on Zenodo\footnote{\url{https://zenodo.org/records/10998726}}. In many cases, we saw morphological similarities between the planet candidate and matching signal, and in other cases, we saw no signal corresponding to the planet candidate, indicating that the candidate may have been an artifact of the particular photometric aperture used for extraction and is therefore also a likely false positive.} On the other hand, sometimes the matched signals had clearly different light curve shapes (implying the ephemerides matched by random chance). In other cases, the match was real, but the pixel level data implied that the known planet candidate was the true source of the signal. From this vetting process, we retained only signals that appeared likely to be matches and where our analysis showed that an otherwise viable planet candidate was likely a false positive. \bedit{We note that although we performed this analysis manually, in the future it will be useful to automate this check using metrics like the transit duration and the ratio of ingress/egress time in addition to the period and transit time for matching.} 

\section{Results}\label{results}

\begin{figure*}[t!]
    \centering
    \includegraphics[width=0.98\textwidth]{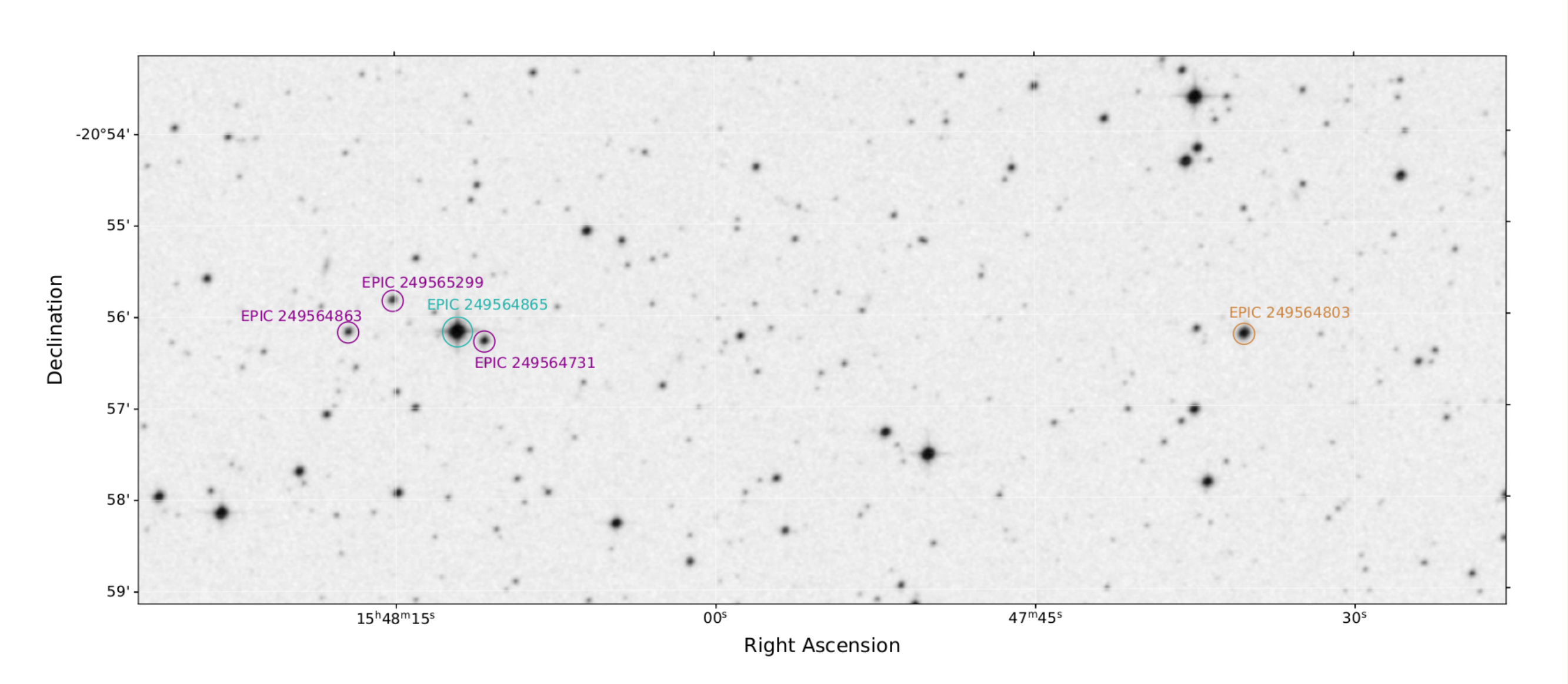}
    \caption{An image of the sky from the Palomar Observatory Sky Survey containing stars that show false positive signals (from the same family of false positives  as shown in Figure \ref{Figure 3}). We identified the source of the family as the bright (\Kepler\ band magnitude $K_p =$10.9) eclipsing binary EPIC 249564865. Several nearby stars (circled in purple) show false positive signals that we suspect are caused by direct PRF overlap. There is one other false positive signal on the more distant star EPIC 249564803. Because this star happens to fall on the same CCD column as the source (EPIC 249564865), this star is likely contaminated via column anomaly.}
    \label{Figure 2}
\end{figure*}

We applied the methodology outlined in the previous section to all K2 campaigns and compiled a list of \bedit{43} false positives among the published K2 planet candidates, including one planet that had previously been statistically validated. 

\subsection{False Positive Planet Candidates}
We list the \bedit{43} false positive planet candidates identified as a result of our analysis in Table \ref{fptable}. In this table, we include a list of the false positive planet candidates (given by the EPIC IDs, and labeled FP EPIC, where FP is an abbreviation for ``False Positive'') and the associated matching eclipsing binary signal (labeled Match EPIC). The table also lists in which K2 campaign the false positive planet candidate was first observed, and the publication reference for the planet candidate. The remaining columns in the table contain information that we used to identify the signal as a false positive, including the distance between the two sources, the periods and transit times of both stars, and the CCD columns of both stars. Most of the false positive appear to be caused by relatively close-by contaminants, indicating they are likely due to direct PRF overlap, but some of them are further away and were observed in close-by CCD columns, indicating column anomaly is likely their source. 

Upon inspection of the false positive list, we noticed that many of the false positives fall into ``families," which are groups of signals all caused by the same parent contaminant. One particularly large family includes a number of different false positive signals in Campaign 5. We show an image of the sky from the Palomar Observatory Sky Survey in Figure \ref{Figure 2} showing the relative locations of the different matching signals, and we show the light curves of each star identified in the family in Figure \ref{Figure 3}. We note the various families of signals that we identified in Table \ref{fptable}.

\begin{figure}[t!]
    \centering
    \includegraphics[width=0.49\textwidth]{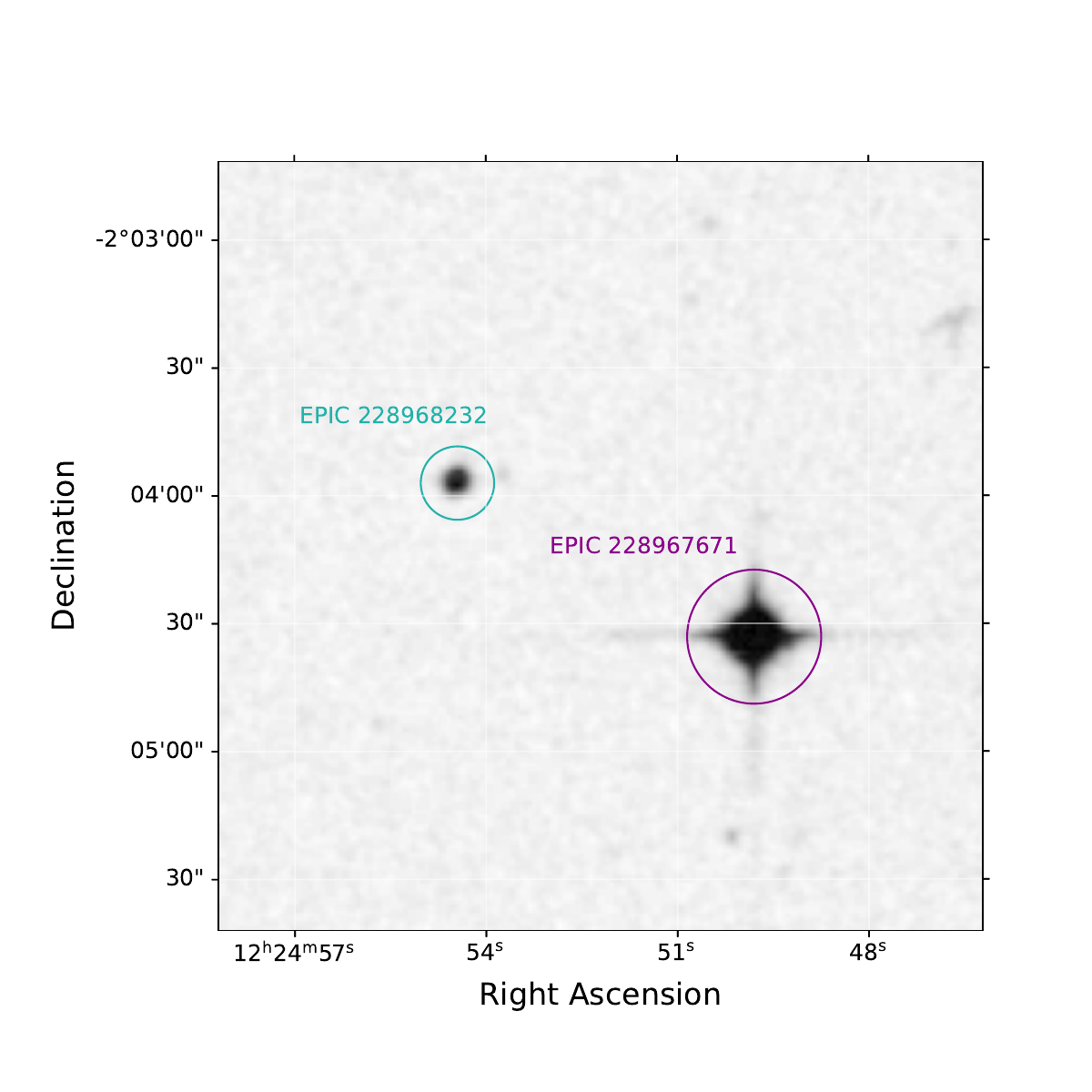}
    \caption{Image of the sky surrounding K2-256 (EPIC 228968232) and its contaminant, the bright (\Kepler-band magnitude = 10.4) eclipsing binary EPIC 228967671. The precisely matched orbital period and transit/eclipse time between the two signals, combined with the two stars' close proximity of these stars is highly suggestive that K2-256 b is a false positive, likely due to direct PRF contamination.}
    \label{Figure 4}
\end{figure}

\subsection{False Positive Validated Planet K2-256 b}

Among the list of \bedit{43} false positive planet candidate was one object considered to be a statistically validated planet: K2-256 b, also known as EPIC 228968232.01. K2-256 b was originally reported and statistically validated by \citet{Livingston2018} as part of a large sample 44 validated planets from K2 Campaign 10. \citet{Livingston2018} used the \vespa\ software \citep{morton12, morton2015} and other various statistical tests to identify and validate planet candidates. The planet was believed to be a super Earth with a radius of approximately $2.63^{+0.25}_{-0.21}$ \rearth\ and a period of $5.52011^{+0.00239}_{-0.00289}$ days, orbiting a relatively faint star (\Kepler\ magnitude of 14.7). \citet{Livingston2018} reported that its host star has an effective temperature of $5219^{+179}_{-136}$ K, and a stellar mass and radius of $0.84^{+0.03}_{-0.04}$ \msun and $0.78\pm0.02$ \rsun, respectively.

Statistically validated planets like K2-256 b are considered overwhelmingly likely to be exoplanets. To be validated, they usually have undergone rigorous vetting in both the \Kepler\ data and follow-up observations, and statistical calculations must have found that any plausible false positive scenarios are very unlikely (typically with confidence of 99\% or 99.9\%). Nevertheless, statistically validated planets have not necessarily been \textit{confirmed} with follow-up observations that re-detect the planet using another independent technique, so it is plausible that some validated planets are in fact false positives. We therefore paid greater attention to K2-256 b to determine whether it indeed is a rare false positive among validated planets. 

\begin{figure*}[t!]
    \centering
    \includegraphics[width=0.98\textwidth]{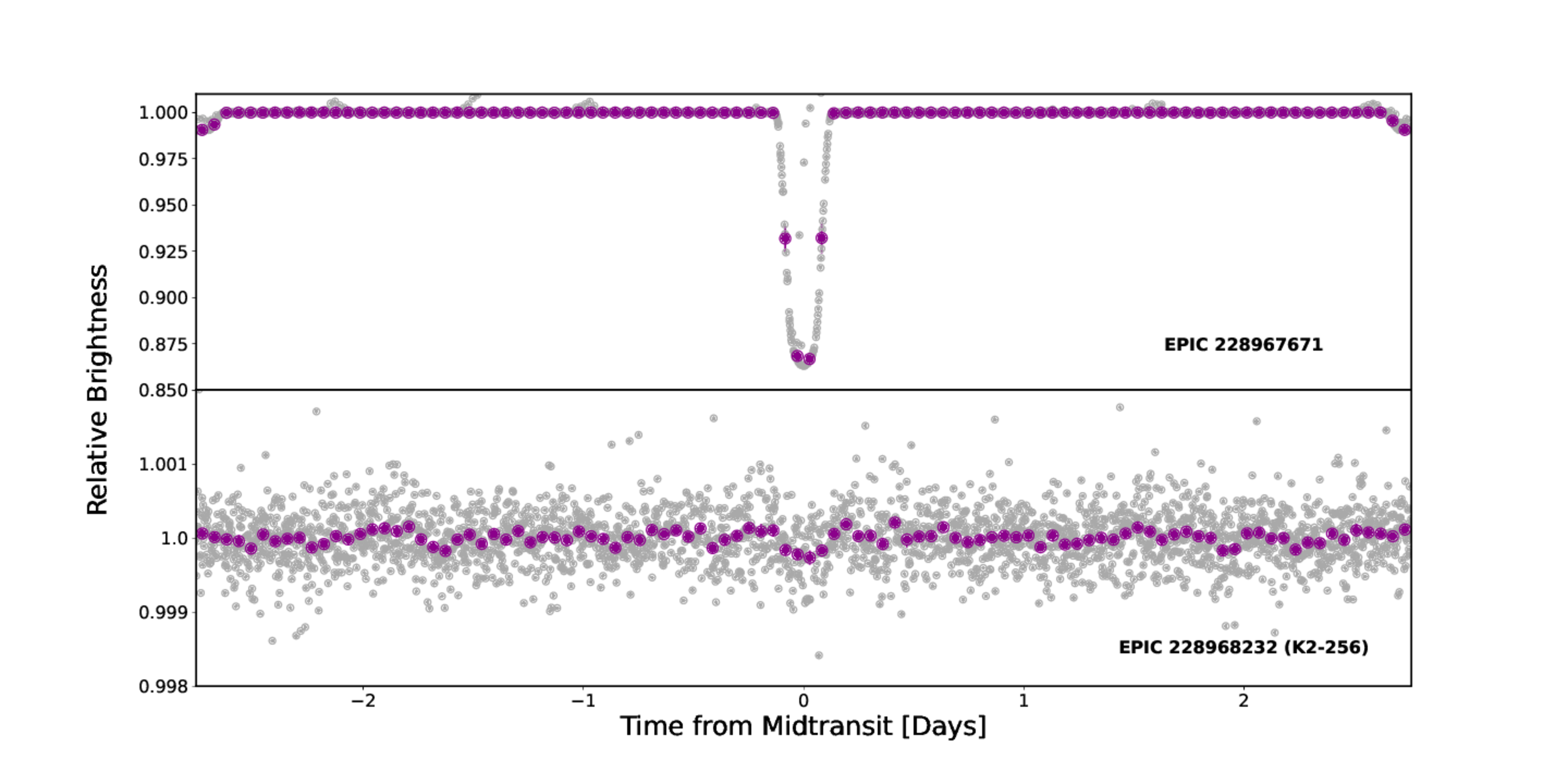}
    \caption{Light curves of the false-positive validated planet host K2-256 (\textit{bottom}) and the nearby bright eclipsing binary star EPIC 228967671 (\textit{top}). The shape and duration of the light dips are consistent in both light curves, indicating that contamination from EPIC 228967671 is the likely source of the transit of K2-256 b. We also note that we see a considerably shallower transit signal for K2-256 b than \citet{Livingston2018}, likely because we used a smaller photometric aperture to extact the light curve (see Figure \ref{Figure 5}).}
    \label{Figure 6}
\end{figure*}

\begin{figure*}[t!]
    \centering
    \includegraphics[width=0.98\textwidth]{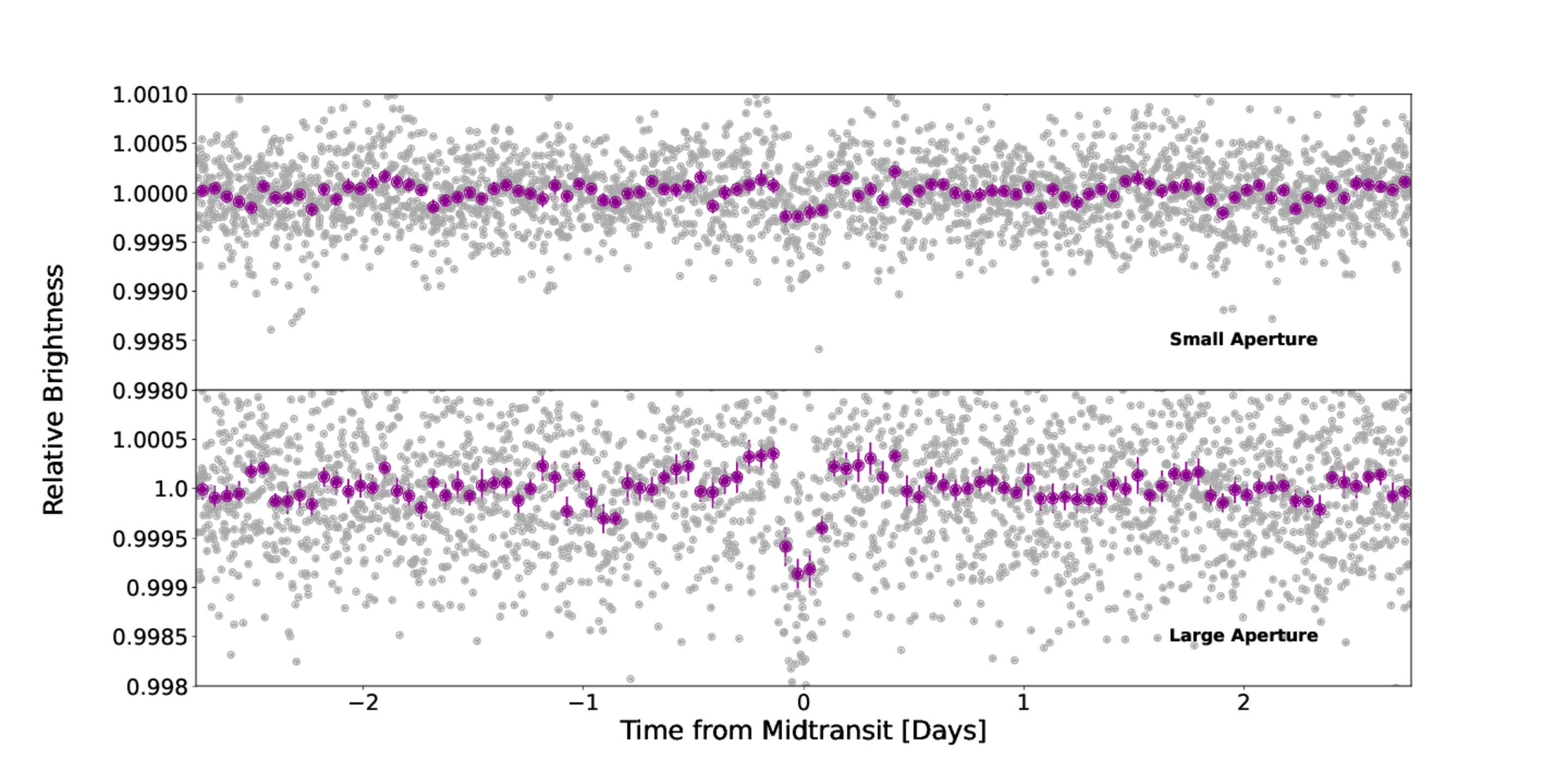}
    \caption{Light curve of the false positive validated planet host K2-256 phase folded on the period of the period and transit time of the false positive signal. We compare two light curves: one extracted with a small photometric aperture window (\textit{top}) and one extracted with a large photometric aperture (\textit{bottom}). We see that the depth of the transit is significantly larger in the light curve extracted with a large aperture -- a telltale sign that the signal is being introduced via scattered background light from the nearby eclipsing binary.}
    \label{Figure 5}
\end{figure*}

Initial tests indicated convincingly that K2-256 b is very likely an ephemeris match false positive.  Figure \ref{Figure 4} shows another image of the sky from the Palmomar Observatory Sky Survey which illustrates the close proximity of the target to its presumed contaminant (EPIC 228967671). With a distance of only about 19.5 pixels (or 78.1 arcseconds, see Table \ref{fptable} on page \pageref{fptable}) between the two, we determined that direct PRF contamination from EPIC 228967671 could plausibly create the signal on K2-256. We also found that the duration and shape of the transits/eclipses seen on K2-256 and EPIC  228967671 were consistent (see Figure \ref{Figure 6}). When plotting the light curve of K2-256 in this test, we also noticed that the depth of the transit of K2-256 b was considerably smaller than what was reported by \citet{Livingston2018}. We therefore plotted light curves of K2-256 extracted from photometric apertures with different sizes in Figure \ref{Figure 5}. We found that larger apertures showed the transit signal with a depth of approximately 0.1\%, while smaller apertures (including the default aperture chosen by our K2 pipeline) showed a much shallower signal (if any signal at all). This is another independent sign that the signal of K2-256 b is likely due to contamination -- apertures with a greater area capture a larger amount of scattered light. Given these multiple, independent lines of reasoning, we conclude that the signal of K2-256 b is in fact originating from the eclipsing binary EPIC 228967671, and is a false positive validated planet. 



\section{Discussion and Conclusions}
\label{discussion}

In this work, we have identified false positive planet candidates from the K2 mission using the technique of ephemeris matching. While this procedure was applied regularly during the \Kepler\ mission to identify false positive \Kepler\ planet candidates, this was not necessarily the case during the K2 mission, so there is value in us performing this analysis on the published K2 candidates. 

We performed the ephemeris matching procedure in a manner very similar to that of \citet{Coughlin2014AJ}, with a few minor modifications needed to adapt the process to the K2 mission. After identifying potential matches and vetting the results, we report a list of \bedit{43} false positive planet candidates, including one planet that had previously been statistically validated: K2-256 b. While it is unusual for validated planets to be false positives, it is not unheard of, especially in situations where the false positive scenario is due to contamination from distant stars \citep[e.g.,][]{Cabrera2017A&A}. That is because codes that perform statistical validation like \vespa\ assume that sources of the signals they validate must be relatively well localized to be near the target star (in order to calculate the likelihood that a background star could contribute the signal in question). The fact that this assumption was broken in the case of K2-256 explains how this false positive signal could have been statistically validated. 

\bedit{
To help mitigate this issue in the future, we note that more modern vetting algorithms such as TRICERATOPS \citep{Giacalone2021}, MOLUSC \citep{Wood2021}, and RAVEN \citep{Hadjigeorghiou2024} explicitly consider stars near the target as the hosts of possible false positives. Integrating these methods along with a systematic analysis of depth-aperture correlations, which revealed K2-256 b as a false positive, could prevent more false positives from being misidentified as planets.
}

The outcome of our work is that we have increased the purity of the K2 planet candidate sample. As of 24 January 2024, the NASA exoplanet archive lists 1800 unique K2 planet candidates; this paper identifies almost 2\% of these as false positives. This means that the surviving candidates are more likely to be real planets as a result of our analysis, enabling astronomers to perform follow-up observations with higher confidence, and the surviving candidates can be included in statistical analysis with less fear of false positive contamination.  

Although we have identified a significant number of false positives among K2 planet candidates, it is plausible that considerably more remain in the K2 planet candidate lists. In order to identify \textit{all} false positives using ephemeris matching, we must know the orbital periods of \textit{all} eclipsing binary stars in the field of view. However, \Kepler\ could not accomplish this because it was only able to download data from a small fraction of its field of view for anaylsis on Earth. Many stars, including potential eclipsing binary contaminants, were simply ignored due to downlink restrictions, and as a result, we cannot use their orbital periods to identify any ephemeris match false positives. However, since the end of the K2 mission, the Transiting Exoplanet Survey Satellite (TESS) has observed much of the ecliptic plane. TESS does not suffer the same downlink restrictions as did \Kepler, so in principle, it can identify all eclipsing binaries in the K2 fields of view. There may be great benefits to performing an ephemeris match on binary stars detected by TESS to identify false positives in the K2 planet candidate samples. 




\acknowledgments
\textbf{Note in manuscript:} During the late stages of manuscript preparation, we became aware of work by \citet{Tarrants2023arXiv}, who independently identified K2-256 as a false positive based on an ephemeris match with binary stars in \textit{Gaia} data. This independent corroboration strengthens our conclusion that this previously validated planet is in fact a false positive. 

We thank Jeff Coughlin, Steve Bryson, and Michelle Kunimoto for useful conversations, and we thank Melinda Soares-Furtado and Juliette Becker for helpful comments on the manuscript. This research has made use of NASA's Astrophysics Data System, the NASA Exoplanet Archive, which is operated by the California Institute of Technology, under contract with the National Aeronautics and Space Administration under the Exoplanet Exploration Program, and the SIMBAD database,
operated at CDS, Strasbourg, France. The National Geographic Society--Palomar Observatory Sky Atlas (POSS-I) was made by the California Institute of Technology with grants from the National Geographic Society. The Oschin Schmidt Telescope is operated by the California Institute of Technology and Palomar Observatory.

This paper includes data collected by the \Kepler\ mission. Funding for the \Kepler\ mission is provided by the NASA Science Mission directorate. Some of the data presented in this paper were obtained from the Mikulski Archive for Space Telescopes (MAST). STScI is operated by the Association of Universities for Research in Astronomy, Inc., under NASA contract NAS5--26555. Support for MAST for non-HST data is provided by the NASA Office of Space Science via grant NNX13AC07G and by other grants and contracts. This work has made use of data from the European Space Agency (ESA) mission {\it Gaia} (\url{https://www.cosmos.esa.int/gaia}), processed by the {\it Gaia} Data Processing and Analysis Consortium (DPAC, \url{https://www.cosmos.esa.int/web/gaia/dpac/consortium}). Funding for the DPAC
has been provided by national institutions, in particular the institutions participating in the {\it Gaia} Multilateral Agreement.\\

Facilities: \facility{\Kepler/K2, Exoplanet Archive, MAST, CDS, ADS}\\

Software: \texttt{K2FOV} \citep{k2fov}, IDL Astronomy Library \citep{idlastronomylibrary},  
          \texttt{matplotlib} \citep{plt},
          \texttt{numpy} \citep{np}, \texttt{astropy} \citep{AstropyCollaboration2022ApJ}

\bibliographystyle{apj}
\bibliography{refs}

\defcitealias{Barros2016}{B16}
\defcitealias{Vanderburg2016}{V16}
\defcitealias{Dattilo2019}{D19}
\defcitealias{Kruse2019}{K19}
\defcitealias{Petigura2018}{P18}
\defcitealias{Pope2016}{P16}
\defcitealias{Crossfield2016}{C16}
\defcitealias{Adams2016}{A16}
\defcitealias{Yu2018}{Y18}
\defcitealias{Zink2021}{Z21}
\defcitealias{Livingston2018}{L18}

\begin{turnpage}
\begin{deluxetable*}{llccccccccccc}\label{fptable}
\tabletypesize{\scriptsize}
\tablefontsize{\scriptsize}
\tablewidth{0pt}
\centering
\tablecaption{False Positive (FP) List \label{fptable}}
\tablehead{\colhead{FP} & \colhead{Match} & \colhead{Camp-} & \colhead{Reference$^1$} & \colhead{Dist} & \colhead{FP Period} & \colhead{Match Period} & \colhead{FP} & \colhead{Match} & \colhead{FP $t_0$} & \colhead{Match $t_0$} & \colhead{Notes$^3$}\\ \colhead{EPIC ID} & \colhead{EPIC ID} & \colhead{aign} & \colhead{} & \colhead{(\arcsec)} & \colhead{(days)} & \colhead{(days)} & \colhead{column} & \colhead{column} & \colhead{(BKJD$^2$)} & \colhead{(BKJD)} & \colhead{}}
\startdata
202710713 & 202685083  & 2 & \citetalias{Barros2016}, \citetalias{Vanderburg2016}, 
\citetalias{Kruse2019}, \citetalias{Dattilo2019} & 346 & 3.326159111 & 3.325935475 & 576.8  & 578.4 & 2062.378215 & 2062.376305 & Also matched 202685801         \\
212435047 & 212409377 & 6 & \citetalias{Kruse2019}, \citetalias{Pope2016}, \citetalias{Barros2016}, \citetalias{Petigura2018}, \citetalias{Dattilo2019} & 2087.7 & 1.115521985 & 1.115570287  & 572.9  & 575 & 2385.443957 & 2385.442877 &  
\\
212432685 & 212409856  & 6 & \citetalias{Pope2016}, \citetalias{Kruse2019}, 
\citetalias{Petigura2018}, \citetalias{Dattilo2019} & 1850.3 & 0.531667431 & 0.531681427  & 956.3  & 955.5 & 2384.997597 & 2384.99579 & 
\\
213408445 & 213338208 & 7 & \citetalias{Kruse2019}, \citetalias{Dattilo2019} & 1003.8 & 2.496485163 & 2.497394389  & 716.8  & 715.9 & 2469.00435  & 2469.003814 & 
\\
246114190 & 246115758  & 12       & \citetalias{Dattilo2019}                                                                                                                                                    & 206.4  & 0.584508963 & 0.584509295  & 606.3  & 605.8     & 2906.081469 & 2906.0769  &                                                              \\
246825406 & 246815557  & 13       & \citetalias{Dattilo2019}                                                                                                                                                    & 396.3  & 0.552902042 & 0.55269643   & 327    & 327.9     & 2989.022511 & 2989.022511 &                                                              \\
247241494 & 247253928  & 13       & \citetalias{Dattilo2019}                                                                                                                                                    & 440.6  & 1.716484506 & 1.717582006  & 1001.6 & 914.1     & 2989.985649 & 2989.957481 & also matched 247253678                                  \\
247576592 & 247576642  & 13       & \citetalias{Dattilo2019}                                                                                                                                                    & 14.8   & 6.969759435 & 6.969797294  & 311.7  & 310.2     & 2992.053616 & 2992.056854 &                                                              \\
249564803 & 249564865  & 15       & \citetalias{Dattilo2019}                                                                                                                                                    & 516.9  & 2.132032054 & 2.131030843  & 322.3  & 321       & 3159.84973  & 3159.862473 & FAMILY 1                                  \\
249564863 & 249564865  & 15       & \citetalias{Dattilo2019}                                                                                                                                                    & 71.9   & 2.130314627 & 2.131030843  & 320.6  & 321       & 3159.859285 & 3159.862473 & FAMILY 1            \\
249565299 & 249564865  & 15       & \citetalias{Dattilo2019}                                                                                                                                                    & 47.2   & 2.130414826 & 2.131030843  & 325.9  & 321       & 3159.871546 & 3159.862473 & FAMILY 1                       \\
211432140 & 211432167  & 16       & \citetalias{Dattilo2019}                                                                                                                                                    & 15.5   & 5.817947423 & 5.817247541  & 575    & 571.1     & 3268.853014 & 3268.85516 & FAMILY 2                       \\
211664909 & 211662047  & 16       & \citetalias{Dattilo2019}                                                                                                                                                    & 146.2  & 0.317621894 & 0.317624179  & 442.5  & 443.1     & 3264.212242 & 3264.212696 &                                                              \\
211784767 & 211785852 & 5 & \citetalias{Barros2016} & 89.8 & 3.5792243 & 3.5792244 & 897.3  & 883.6 & 2869.1174   & 2869.1191 & 
\\
211808055 & 211807843  & 5 & \citetalias{Pope2016}, 
\citetalias{Barros2016} & 20.4 & 3.382079 & 3.3820746 & 130.5 & 126.1 & 2867.6547 & 2867.6553 & 
\\
201270176 & 201270464  & 1        & \citetalias{Barros2016}, \citetalias{Dattilo2019}                                                                                                               & 16.4   & 1.57771549  & 1.578162319  & 932    & 932.2     & 1985.327    & 1977.426053 &                                                              \\
201270464 & 201270176  & 1        & \citetalias{Barros2016}, \citetalias{Vanderburg2016}, \citetalias{Dattilo2019}                                                                      & 16     & 1.57771549  & 1.578198138  & 932.2  & 932       & 1985.328    & 1977.425374 &                                                              \\
203518244 & 203485624  & 2        & \citetalias{Vanderburg2016}                                                                                                                                                 & 398.7  & 0.8411257   & 0.841211423  & 347.4  & 346.3     & 2061.581    & 2061.581281 &                                                              \\
204676499 & 204676803  & 2        & \citetalias{Barros2016}                                                                                                                                                     & 21     & 1.5271792   & 1.52729779   & 925.1  & 930.3     & 2064.469    & 2061.414287 & also matched 204676841                                  \\
205703094 & 205703649  & 2        & \citetalias{Crossfield2016}                                                                                                                                                   & 22.3   & 8.1191      & 4.058421328  & 46.5   & 50.1      & 2062.817    & 2062.83691 & 3 planet system                                              \\
205962305 & 205962680  & 3        & \citetalias{Kruse2019}                                                                                                                                                      & 67.1   & 1.88282     & 1.88279906   & 1035.1 & 1030.4    & 2144.839    & 2146.723148 &                                                              \\
211432922 & 211432167  & 5        & \citetalias{Kruse2019}                                                                                                                                                      & 50.5   & 5.81861     & 5.817716589  & 1009.8 & 1003      & 2308.923    & 2308.929664 & FAMILY 2                                                             \\
211685045 & 211685048  & 5        & \citetalias{Adams2016}                                                                                                                                                      & 10.8   & 0.769057    & 0.769144365  & 108.1  & 108.1     & 2368.384    & 2307.622276 &                                                              \\
211834065 & 211834405  & 5        & \citetalias{Pope2016}                                                                                                                                                       & 596.4  & 10.545      & 10.54236357  & 185.1  & 184.9     & 2309.426    & 2309.431152 &                                                              \\
211914889 & 211915147  & 5        & \citetalias{Yu2018}                                                                                                                                                         & 303.5  & 1.81042     & 1.810855435  & 843.7  & 767.4     & 3263.481    & 2309.196795 &                                                              \\
211914960 & 211915147  & 5        & \citetalias{Yu2018}                                                                                                                                                         & 750.4  & 1.810834    & 1.810855435  & 956.2  & 767.4     & 3263.469    & 2309.196795 &                                                              \\
211995325 & 211995966  & 5        & \citetalias{Adams2016}                                                                                                                                                      & 37.4   & 0.279258    & 0.279261028  & 269.2  & 268.2     & 2321.452    & 2307.768553 &                                                              \\
212024672 & 212024647  & 5        & \citetalias{Yu2018}                                                                                                                                                         & 475.3  & 3.697161    & 3.696849632  & 245.2  & 245.3     & 3262.873    & 2309.097988 &                                                              \\
212351026 & 212351048  & 6        & \citetalias{Barros2016}, \citetalias{Pope2016}                                                                                                                  & 61.2   & 2.55068936  & 2.548684118  & 714.8  & 730.2     & 2392.785    & 2385.160991 & FAMILY 3                       \\
212351405 & 212349118  & 6        & \citetalias{Barros2016}, \citetalias{Pope2016}                                                                                                                  & 213.4  & 2.549       & 2.549282799  & 736.7  & 732.6     & 2385.158    & 2385.14609 & FAMILY 3                       \\
220434612 & 220434992  & 8        & \citetalias{Zink2021}                                                                                                                                                       & 26.8   & 2.010607    & 2.010046489  & 971.7  & 977.6     & 2560.924    & 2560.927644 &                                                              \\
201164625 & 201166041  & 10       & \citetalias{Livingston2018}                                                                                                                                                 & 120.2  & 2.71189     & 2.711471052  & 131.5  & 103.4     & 2750.136    & 2750.139122 &                                                              \\
228968232 & 228967671  & 10       & \citetalias{Livingston2018}                                                                                                                                                 & 78.1   & 5.52011     & 5.518602409  & 333    & 317.2     & 2753.525    & 2753.515935 & K2-256 b                                  \\
211616939 & 211620138  & 16       & \citetalias{Yu2018}                                                                                                                                                         & 162.1  & 1.8554      & 1.855512984  & 812.8  & 813.9     & 3262.619    & 3264.473398 & FAMILY 4 \\
211619805 & 211620138  & 16       & \citetalias{Yu2018}                                                                                                                                                         & 27.5   & 1.855827    & 1.855512984  & 819.5  & 813.9     & 3262.61     & 3264.473398 & FAMILY 4                       \\
211620138 & 211618939  & 16       & \citetalias{Yu2018}                                                                                                                                                         & 66.3   & 1.855421    & 1.85537045   & 813.9  & 809.5     & 3262.619    & 3264.476333 & FAMILY 4            \\
211663688 & 211620138  & 16       & \citetalias{Yu2018}                                                                                                                                                         & 2182.6 & 1.855516    & 1.855512984  & 816.5  & 813.9     & 3262.615    & 3264.473398 & FAMILY 4            \\
211863149 & 211839462  & 16       & \citetalias{Yu2018}                                                                                                                                                         & 1172.9 & 2.61299     & 2.613052395  & 94.5   & 94.8      & 3264.303    & 3264.30344 & also matched 211839430                                  \\
211964332 & 211964025  & 16       & \citetalias{Yu2018}                                                                                                                                                         & 1298.2 & 7.220537    & 7.220257782  & 145.3  & 144.7     & 3266.504    & 3266.503813 & also matched 211964001                                  \\
\bedit{211972627} &  211972681 & 16       & \citetalias{Yu2018}                                                                                                                                                         & 157.7  &   1.092999714  & 1.092673  &  23.7   & 25.3     &  3264.546252  & 3263.463  & Source is actually 211972837                                  \\
\bedit{211972681} & 211972627  & 16       & \citetalias{Yu2018}                                                                                                                                                         & 157.7  & 1.092673    & 1.092999714  & 25.3   & 23.7      & 3263.463    & 3264.546252 & Source is actually 211972837                                  \\
212207368 & 212223307  & 16       & \citetalias{Yu2018}                                                                                                                                                         & 2014.4 & 1.190316    & 1.19037742   & 915.7  & 918.7     & 3263.292    & 3264.482842 & FAMILY 5                                  \\
212223307 & 212223632 & 16 & \citetalias{Yu2018} & 70.5   & 1.190354    & 1.190180933  & 918.8  & 931.1 & 3263.292 & 3264.488926 & FAMILY 5\\
\enddata
\tablenotetext{1}{
A16 : \citet{Adams2016}, 
B16 : \citet{Barros2016}, 
C16 : \citet{Crossfield2016}, 
D19 : \citet{Dattilo2019}, 
K19 : \citet{Kruse2019}, 
L18 : \citet{Livingston2018}, \\
P18 : \citet{Petigura2018}, 
P16 : \citet{Pope2016}, 
V16 : \citet{Vanderburg2016}, 
Y18 : \citet{Yu2018}, 
Z21 : \citet{Zink2021}\\
}

\tablenotetext{2}{BKJD is the standard \Kepler\ mission time standard, equivalent to BJD$_{\rm TDB} - 2454833$\\}
\tablenotetext{3}{Sometimes eclipsing binaries can contaminate multiple other stars and create a family of false positives. We list the members of several false positive families here:\\
FAMILY 1: 249564865, 249564803, 249564863, 249565299, 249564731 \\
FAMILY 2: 211432167, 211432140, 211432176, 211432922 \\
FAMILY 3: 212351048, 212349118, 212351026, 212351405, 212351542, 212351335 \\
FAMILY 4: 211620138, 211618939, 211616939, 211619805, 211663688, 211618022, 211623771 \\
FAMILY 5: 212223307, 212223632, 212207368, 212222875, 212222443 \\
}
\end{deluxetable*}
\end{turnpage}

\end{document}